\documentclass[superscriptaddress,preprint,amsmath,amssymb,aps,pre]{revtex4}
\usepackage{graphicx} 

\begin{document}
\title{Making public reputation out of private assessments}
\author{Youngsuk Mun}
\thanks{These two authors contributed equally.}
\affiliation{Department of Physics, Pukyong National University, Busan 48513, Korea}
\author{Quang Anh Le}
\thanks{These two authors contributed equally.}
\affiliation{Industry-University Cooperation Foundation, Pukyong National University, Busan 48513, Korea}
\author{Seung Ki Baek}
\email{seungki@pknu.ac.kr}
\affiliation{Department of Scientific Computing, Pukyong National University, Busan 48513, Korea}
\date{\today}

\begin{abstract}
Reputation is not just a simple opinion that an individual has
about another but a social construct that emerges through communication.
Despite the huge importance in coordinating human behavior,
such a communicative aspect has remained relatively unexplored in the field of indirect reciprocity.
In this work, we bridge the gap between private assessment and public reputation: We begin by clarifying what we mean by reputation and argue that the formation of reputation can be modeled by a bi-stochastic matrix, provided that both assessment and behavior are regarded as continuous variables. By choosing bi-stochastic matrices that represent averaging processes, we show that only four norms among the leading eight, which judge a good person's cooperation toward a bad one as good, will keep cooperation asymptotically or neutrally stable against assessment error in a homogeneous society where every member has adopted the same norm. However, when one of those four norms is used by the resident population, the opinion averaging process allows neutral invasion of mutant norms with small differences in the assessment rule.
Our approach provides a theoretical framework for describing the formation of reputation in mathematical terms.
\end{abstract}
\maketitle

\section{Introduction}

The Cambridge Dictionary defines reputation as ``the opinion that people have
about someone or something based on their behaviour or character in the past.''
Quite often when we cannot objectively measure people's opinion,
what we mean by someone's reputation will actually correspond to our opinion
about people's opinion about the person~\cite{origgi2017reputation}.
The point is that reputation is not just a simple individual opinion but something that emerges on the societal level.

Despite the huge importance in human interactions~\cite{dunbar2004gossip},
the communicative aspect of reputation formation has remained relatively unexplored in the field of indirect reciprocity~\cite{nowak2005evolution,xia2023reputation,santos2023consensus}:
The first mathematical treatment simply assumed the existence of `image scores' known to every player~\cite{nowak1998evolution}.
The indirect-observation model provides a mechanism for synchronizing private assessments~\cite{ohtsuki2004should}, in which only one individual plays the role of an observer at each given moment and spreads his or her own assessment to the whole population.
This model could be close to the truth when news was rare and usually delivered by small parties of travelers~\cite{burke1985day}, but such a narrow one-way communication channel is not always realistic.
At the other extreme, one may study the dynamics when no synchronizing mechanism exists among different opinions. Overall, this line of research indicates that cooperation becomes more vulnerable to noise~\cite{hilbe2018indirect,schmid2021evolution,schmid2023quantitative,fujimoto2023evolutionary,murase2024computational}, but it should be admitted that the complete lack of communication in this scenario is another type of idealization.

Beyond these two extremes, some ideas of opinion exchange have been put forward (see, e.g.,
Refs.~\cite{ohtsuki2009indirect,takacs2021networks,kessinger2023evolution,kawakatsu2024mechanistic,murase2024indirect}), but
we believe that the process of reputation formation deserves more systematic investigation.
As the first step, we wish to clarify what we mean by reputation, consider a simple mechanism according to which the convergence of reputation can be proved, and see how this additional process affects the
dynamics of assessment and behavior toward others.
We will be using the continuous model of indirect reciprocity, which regards both assessment and behavior as continuous variables~\cite{lee2021local,lee2022second,mun2023second}. In addition to mathematical convenience, this approach provides a way to overcome the sharp dichotomy between good and bad or between cooperation and defection (see also Ref.~\cite{schmid2023quantitative} for comparison), together with a qualitative agreement with the original discrete model of indirect reciprocity~\cite{fujimoto2024leader}.

This work is organized as follows: In the next section, we explain how we understand reputation and propose its mathematical formulation, which is followed by a short review for the continuous model of indirect reciprocity. In Sec.~\ref{sec:result}, we check how the formation of public reputation based on private assessments affects the stability against error or mutation. Section~\ref{sec:discuss} discusses the findings and summarizes this work.


\section{Model}
\label{sec:model}

\subsection{Reputation formation}

By reputation we mean a publicly shared opinion resulting from aggregation of individual assessments. Let us imagine an ideal reputation-forming process $F$ and define $m_{ij}$ as an individual $i$'s assessment of another individual $j$. By applying $F$ to the set of bare assessments $\left\{ m_{ij}^{(0)} \right\}$, we obtain higher-level assessments:
\begin{equation}
\left\{ m_{ij}^{(1)} \right\} = F \left( \left\{ m_{ij}^{(0)} \right\} \right),
\end{equation}
which generalizes to
\begin{equation}
\left\{ m_{ij}^{(n)} \right\} = F^n \left( \left\{ m_{ij}^{(0)} \right\} \right),
\end{equation}
where $n$ is the level of assessments, which may also be interpreted as time if we regard reputation formation as a temporal dynamics. If the process $F$ is applied infinitely many times, we should obtain
\begin{equation}
\left\{ m_{ij}^{(\infty)} \right\} = \lim_{n \to \infty} F^n \left( \left\{ m_{ij}^{(0)} \right\} \right),
\end{equation}
in which everyone agrees on everyone's goodness, i.e., $m_{ij}^{(\infty)} = m_{kj}^{(\infty)}$ for any $i$, $j$, and $k$. Noting that the first index is irrelevant, we may just call this quantity $m_j^{(\infty)}$ and regard it as individual $j$'s reputation.

One of the simplest ways to implement $F$ is found in linear algebra: Let us consider a doubly stochastic matrix $W = \left\{ w_{ik} \right\}$ that describes an irreducible aperiodic finite Markov chain with $w_{ik}\ge 0$ and $\sum_i w_{ik} = \sum_k w_{ik} = 1$. If the process $F$ is implemented as
\begin{equation}
    m^{(n+1)}_{ij} = \sum_k w_{ik} m^{(n)}_{kj},
\end{equation}
the Perron-Frobenius theorem assures that an $N$-dimensional vector
\begin{equation}
    \vec{m}_j^{(n)} \equiv \left( m_{1j}^{(n)}, m_{2j}^{(n)}, \ldots, m_{Nj}^{(n)} \right)^\intercal
\end{equation}
will almost always become parallel to the principal eigenvector as $n\to \infty$, regardless of the initial direction of $\vec{m}_j^{(0)}$. Moreover, the elements of the principal eigenvector are uniform, i.e., $m_{ij}^{(\infty)} = m_{kj}^{(\infty)}$ for any $i$, $j$, and $k$, if and only if $W$ is doubly stochastic.
This is exactly the desired property for reputation.
In fact, as a stochastic matrix, $W$ preserves the total sum of elements in such a way that $\sum_k m_{ki}^{(0)} = \sum_k m_{ki}^{(\infty)} = N m_{i}^{(\infty)}$. Therefore, an individual $i$'s reputation is
fully determined by $m_i^{(\infty)} = N^{-1} \sum_k m_{ki}^{(0)}$, the average assessment received at the beginning.

\subsection{Normative dynamics}

By normative dynamics, we mean the dynamics governed by a social norm, which tells an individual how to assess others' behavior and how to behave toward them based on the assessments. Note that it can work on the level of individual assessment without requiring public reputation in the above sense.
We will consider a continuous model of indirect
reciprocity by regarding both assessment and behavior as continuous
variables~\cite{lee2021local,lee2022second,mun2023second}:
Assume that we have a population of $N$ individuals. Each individual $k$ has a
real-valued opinion about another individual $i$, which we call $m_{ki} \in
[0,1]$, by assessing $i$'s behavior toward someone else. A higher value of
$m_{ki}$ means a more positive opinion that $k$ has about $i$.
Mathematically speaking, each individual $k$ has an assessment rule $\alpha_k
(m_{ki}, \beta_i, m_{kj})$ to determine the opinion about $i$, which
depends on $k$'s original assessment of $i$, $i$'s behavior to $j$, and $k$'s assessment
of $j$. Likewise, each individual $i$ has a behavioral rule $\beta_i
\left(m_{ii}, m_{ij} \right) \in [0,1]$ to determine the degree of cooperation
toward $j$ depending on the self-image $m_{ii}$ as well as his or her own assessment
of $j$.
We consider the donation game parameterized by the benefit $b$ and cost $c$ of cooperation with $b>c>0$: When two individuals $i$ and $j$ are picked as a donor and a recipient, respectively, the donor $i$ cooperates by donating $b \beta_i(m_{ii}, m_{ij})$ to the recipient $j$ at the cost of $c \beta_i(m_{ii}, m_{ij})$.

If the observer $k$'s opinion $m_{ki}$ is updated when the interaction between $i$ and $j$ is
observed with probability $q>0$, the dynamics can be
formulated as follows:
\begin{equation}
m_{ki}^{t+1} = (1-q) m_{ki}^t + \frac{q}{N-1} \sum_{j \neq i} \alpha_k \left[
m_{ki}^t, \beta_i \left(m_{ii}^t, m_{ij}^t \right), m_{kj}^t \right],
\label{eq:update}
\end{equation}
where the superscripts $t$ and $t+1$ mean time steps. On the right-hand side of
Eq.~\eqref{eq:update}, we have taken average over
$j$ because $i$ can meet anyone else with equal probability. In this formulation, the observation probability $q$ only rescales the overall time scale.

\begin{table}
    \caption{Continuous versions of the leading eight~\cite{ohtsuki2004should,ohtsuki2006leading}.
    We assign numerical values $1$ and $0$ to good and bad (or cooperation and
    defection, depending on the context), respectively.
    By substituting the boundary values $1$ and $0$ into $x$, $y$, and $z$,
    these functional descriptions recover all the original prescriptions of the
    leading eight as given by Ref.~\cite{ohtsuki2004should}.
    }
    \centering
    \begin{tabular}{l|c|c}\hline
        Norm & $\alpha(x,y,z)$ & $\beta(x,y)$ \\\hline\hline
        L1 & $x+y-xy-xz+xyz$ & $-x+xy+1$\\
        L2 & $x+y-2xy-xz+2xyz$ & $-x+xy+1$\\
        L3 & $yz - z + 1$ & $y$\\
        L4 & $-y-z+xy+2yz-xyz+1$ & $y$\\
        L5 & $-z-xy+yz+xyz+1$ & $y$\\
        L6 & $-y-z+2yz+1$ & $y$\\
        L7 & $x-xz+yz$ & $y$\\
        L8 & $x-xy-xz+yz+xyz$ & $y$\\\hline
    \end{tabular}
    \label{tab:cont}
\end{table}

Suppose that everyone uses a common norm among the leading eight~\cite{ohtsuki2004should,ohtsuki2006leading}, i.e.,
$\alpha_k = \alpha$ and $\beta_i = \beta$ as in Table~\ref{tab:cont}.
All the eight norms in Table~\ref{tab:cont} have full cooperation as a trivial fixed point, at which
$m_{ki}=1$ for every $k$ and $i$. Assume that $m_{ki} = m^\ast$ is a general fixed point, at which the following equality holds:
\begin{equation}
m^\ast = \alpha \left[ m^\ast, \beta(m^\ast, m^\ast), m^\ast \right],
\label{eq:single}
\end{equation}
independently of $q$ and $N$.
Then, we add perturbation here by considering error, whose probability is low enough that the dynamics of Eq.~\eqref{eq:update} can be assumed to be valid while approaching a stationary state.
By defining $\epsilon_{ki} \equiv m^\ast - m_{ki}$ with $|\epsilon_{ki}| \ll 1$ and expanding Eq.~\eqref{eq:update} to the first order of these small
parameters, we obtain
\begin{equation}
\epsilon_{ki}^{t+1} = (1-q) \epsilon_{ki}^t + qA_x \epsilon_{ki}^t
+ qA_y B_x \epsilon_{ii}^t + \frac{q}{N-1}
\sum_{j \neq i} [A_y B_y \epsilon_{ij}^t + A_z \epsilon_{kj}^t],
\label{eq:disagree}
\end{equation}
where
\begin{subequations}
\begin{align}
A_x &\equiv \left.\partial_x \alpha(x,y,z) \right|_{(m^\ast,m^\ast,m^\ast)}\\
A_y &\equiv \left.\partial_y \alpha(x,y,z) \right|_{(m^\ast,m^\ast,m^\ast)}\\
A_z &\equiv \left.\partial_z \alpha(x,y,z) \right|_{(m^\ast,m^\ast,m^\ast)}\\
B_x &\equiv \left.\partial_x \beta(x,y) \right|_{(m^\ast,m^\ast)}\\
B_y &\equiv \left.\partial_y \beta(x,y) \right|_{(m^\ast,m^\ast)}.
\end{align}
\label{eq:derivatives}
\end{subequations}
Symbolically, we may write it as
\begin{equation}
    \vec{\epsilon}(t+1) = \mathcal{Q} \vec{\epsilon} (t)
    \label{eq:Q}
\end{equation}
by introducing an $N^2$-dimensional vector
\begin{equation}
    \vec{\epsilon}(t) \equiv \left(\epsilon_{11}^t, \ldots,
\epsilon_{NN}^t\right)^\intercal
\label{eq:epsilon_vector}
\end{equation}
and an $N^2 \times N^2$ matrix $\mathcal{Q}$.
The eigenvalue structure of $\mathcal{Q}$ is obtained as follows~\cite{lee2021local}:
\begin{subequations}
\begin{align}
\Lambda_1^{[N^2-2N+1]} &= (1-q)+q\left(A_x- \frac{1}{N-1}A_z \right)\\
\Lambda_2^{[N-1]} &= (1-q)+q(A_x+A_z)\\
\Lambda_3^{[N-1]} &= (1-q)+q\left(A_x-\frac{1}{N-1}A_z+A_y B_x
-\frac{1}{N-1}A_y B_y \right)\\
\Lambda_4^{[1]} &= (1-q)+q(A_x + A_z + A_y B_x + A_y B_y),
\end{align}
\label{eq:eig}
\end{subequations}
where each superscript on the left-hand side
means the multiplicity of the eigenvalue.
The largest eigenvalue is $\Lambda_4^{[1]}$ when all the derivatives are
positive, and the corresponding eigenvector is
\begin{equation}
\vec{V} = (1,1,\ldots,1)^\intercal
\label{eq:eigenvector}
\end{equation}
up to a nonzero proportionality constant~\cite{lee2022second}. Another set of eigenvectors associated with $\Lambda_3^{[N-1]}$ are obtained as follows:
\begin{equation}
\begin{array}{c}
(\underbrace{1,-1,0,\ldots,0}_{N},\underbrace{1,-1,0,\ldots,0}_{N},\ldots, \underbrace{1,-1,0,\ldots,0}_{N}),\\
(\underbrace{1,0,-1,\ldots,0}_{N},\underbrace{1,0,-1,\ldots,0}_{N},\ldots, \underbrace{1,0,-1,\ldots,0}_{N}),\\
\vdots\\
(\underbrace{1,0,0,\ldots,-1}_{N},\underbrace{1,0,0,\ldots,-1}_{N},\ldots, \underbrace{1,0,0,\ldots,-1}_{N}).
\label{eq:eigenvector_lambda3}
\end{array}
\end{equation}

Our numerical simulations of Eq.~\eqref{eq:update} show that $m_{ki}$ approaches
a certain value
$m^\ast$ for every $k$ and $i$ as $t$ grows (not shown), as postulated by Eq.~\eqref{eq:single}, when the initial matrix elements
are sampled uniformly from the unit interval. By solving Eq.~\eqref{eq:single}, we obtain fixed points as follows:
\begin{equation}
m^\ast = \left\{
\begin{array}{ll}
1 & \text{~for L1, L3, and L4}\\
\frac{1}{3} (1+\kappa-2\kappa^{-1}) \approx 0.647799 & \text{~for L2}\\
\varphi^{-1} \approx 0.618034 & \text{~for L5}\\
1/2 & \text{~for L6}\\
\text{underdetermined} & \text{~for L7}\\
0 & \text{~for L8}
\end{array}
\right.
\label{eq:m_ast}
\end{equation}
where $\kappa \equiv (13+3\sqrt{33})^{1/3}/2^{2/3}$, and $\varphi \equiv
(1+\sqrt{5})/2$ is the golden ratio.

\begin{table}
    \caption{First-order derivatives of the continuous leading eight at
    $m_{ki}=m^\ast$ for every $k$ and $i$ [Eq.~\eqref{eq:m_ast}]. We have excluded L1, L3, L4, and L7
    from this table
    because their stability analyses have already been given in
    Ref.~\cite{lee2022second}. For L2, we only show approximate numerical values
    for the sake of brevity.
    }
    \centering
    \begin{tabular}{c|ccccc}\hline
        Norm & $A_x^\ast$ & $A_y^\ast$ & $A_z^\ast$ & $B_x^\ast$ & $B_y^\ast$
        \\\hline\hline
        L2 & -0.10411 & 0.543689 & 0.191488 & -0.352201 & 0.647799\\
        L5 & $1-2\varphi^{-1}$ & $\varphi^{-2}$ & 0 & 0 & 1\\
        L6 & 0 & 0 & 0 & 0 & 1\\
        L8 & 1 & 0 & 0 & 0 & 1\\\hline
    \end{tabular}
    \label{tab:first}
\end{table}
The trivial fixed point $m^\ast=1$ has already been analyzed in Ref.~\cite{lee2022second}. It is asymptotically stable for L1, L3, and L4, and neutrally stable for L7.
Table~\ref{tab:first} shows the values of these derivatives for the rest of the leading eight, i.e., L2, L5, L6, and
L8. Note that some of them take negative values. The eigenvalue structure is still given by Eq.~\eqref{eq:eig}, but care is
needed because the largest eigenvalue may not be necessarily $\Lambda_4^{[1]}$
when some derivatives are negative.
By carefully checking the magnitudes of eigenvalues, we see that
the fixed points in Eq.~\eqref{eq:m_ast} are stable for L2, and L5, respectively. Although the results are indeterminate for L6 and L8,
we can say that that it is also stable based on numerical data.

\section{Results}
\label{sec:result}

\subsection{Error recovery}

\subsubsection{Uniform weights}
To be as simple as possible, let us assume that the weight assigned to every player is uniform, i.e., $w_{ik} = 1/N$. We furthermore assume that $F$ has an equal time scale to that of the normative dynamics
[Eq.~\eqref{eq:update}] so that $W$ and $\mathcal{Q}$ are alternately
acting on the system.
Before proceeding, however, we should
recall that $\mathcal{Q}$ is an $N^2 \times N^2$ matrix, whereas $W$ is defined as an $N \times N$ matrix: We need to construct an $N^2 \times N^2$ matrix $\tilde{W}$ acting on $\vec{\epsilon}^{(n)} = \left(\epsilon_{11}^{(n)}, \ldots, \epsilon_{NN}^{(n)} \right)^\intercal$ based on $W$. For example, when $N=2$, we have
\begin{equation}
    W = \begin{pmatrix}
        \frac{1}{2} & \frac{1}{2}\\
        \frac{1}{2} & \frac{1}{2}
    \end{pmatrix}
\end{equation}
acting on either
$\vec{m}_1^{(n)} \equiv \left( m_{11}^{(n)}, m_{21}^{(n)}\right)^\intercal$ or
$\vec{m}_2^{(n)} \equiv \left( m_{12}^{(n)}, m_{22}^{(n)}\right)^\intercal$. The corresponding $4\times 4$ matrix equation for elevating $\vec{\epsilon}^{(n)}$ to $\vec{\epsilon}^{(n+1)}$ must be
\begin{equation}
\begin{pmatrix}
\epsilon_{11}^{(n+1)}\\
\epsilon_{12}^{(n+1)}\\
\epsilon_{21}^{(n+1)}\\
\epsilon_{22}^{(n+1)}
\end{pmatrix}
=\begin{pmatrix}
\frac{1}{2} & 0 & \frac{1}{2} & 0\\
0 & \frac{1}{2} & 0 & \frac{1}{2}\\
\frac{1}{2} & 0 & \frac{1}{2} & 0\\
0 & \frac{1}{2} & 0 & \frac{1}{2}
\end{pmatrix}
\begin{pmatrix}
\epsilon_{11}^{(n)}\\
\epsilon_{12}^{(n)}\\
\epsilon_{21}^{(n)}\\
\epsilon_{22}^{(n)}
\end{pmatrix},
\end{equation}
which can be rewritten as
$\vec{\epsilon}^{(n+1)} = \tilde{W} \vec{\epsilon}^{(n)}$.
The stability of the combined dynamics can be analyzed by
looking at the eigenvalue structure of $\mathcal{Q} \tilde{W}$, but
it is instructive to begin by checking the eigenvalue structure of $\tilde{W}$ alone. It has two distinct eigenvalues,
\begin{subequations}
\begin{align}
\mu_1^{[N]} &= 1\\
\mu_2^{[N(N-1)]} &= 0,
\end{align}
\end{subequations}
and the eigenvectors associated with $\mu_1^{[N]}=1$ are expressed as
\begin{equation}
\begin{array}{ccc}
\vec{u}_1 & = &(\underbrace{1,0,0,\ldots,0}_{N},\underbrace{1,0,0,\ldots,0}_{N},\ldots, \underbrace{1,0,0,\ldots,0}_{N}),\\
\vec{u}_2 & = &(\underbrace{0,1,0,\ldots,0}_{N},\underbrace{0,1,0,\ldots,0}_{N},\ldots, \underbrace{0,1,0,\ldots,0}_{N}),\\
 &  &\vdots\\
\vec{u}_N & = &(\underbrace{0,0,0,\ldots,1}_{N},\underbrace{0,0,0,\ldots,1}_{N},\ldots, \underbrace{0,0,0,\ldots,1}_{N}).
\end{array}
\label{eq:uk}
\end{equation}
The vectors in Eq.~\eqref{eq:uk} span an $N$-dimensional subspace, which contains both Eqs.~\eqref{eq:eigenvector} and \eqref{eq:eigenvector_lambda3}. The remaining $N(N-1)$-dimensional subspace contains the other set of eigenvectors associated with $\mu_2^{[N(N-1)]} = 0$.
We thus see that $\Lambda_4^{[1]}$ and $\Lambda_3^{[N-1]}$ must still be the largest eigenvalues of $\mathcal{Q} \tilde{W}$, whereas all its
other eigenvalues are zeros. In short, the additional process of reputation formation
effectively reduces the dynamics to a two-level Markov system, accelerating
the convergence to $m^\ast$.

\subsubsection{Nonuniform weights}
We have assumed uniform weights, but it is entirely plausible that one tends to put a different weight on his or her own opinion that any others'. For this reason, let us generalize the above weight matrix $W$ to
\begin{equation}
W' = \theta I + (1-\theta) W
\end{equation}
with $0 \le \theta \le 1$,
where $I$ is an $N \times N$ identity matrix and $W$ is the matrix of uniform weights $w_{ij} = 1/N$. The elements of $W'$ can thus be expressed by
\begin{equation}
w'_{ik} = \left\{
\begin{array}{ll}
\theta + (1-\theta)/N & \text{if~} i=k\\
(1-\theta)/N & \text{otherwise}.
\end{array}
\right.
\end{equation}
As above, we construct an $N^2 \times N^2$ matrix $\tilde{W}'$ based on $W'$ and find the eigenvalue structure of $\tilde{W}'$. The results are given as follows:
\begin{subequations}
\begin{align}
{\mu'}_1^{[N]} &= 1\\
{\mu'}_2^{[N(N-1)]} &= \theta,
\end{align}
\end{subequations}
and the eigenvectors associated with ${\mu'}_1^{[N]}$ are the same as in Eq.~\eqref{eq:uk}. Therefore, compared to Eq.~\eqref{eq:eig}, the first two eigenvalues of $\mathcal{Q} \tilde{W}'$ transform to $\theta \Lambda_1^{[N^2-2N+1]}$ and $\theta \Lambda_2^{[N-1]}$, respectively, whereas $\Lambda_3^{[N-1]}$ and $\Lambda_4^{[1]}$ are left unchanged.

If reputation forms in a shorter time scale than that of the normative dynamics, it will be more appropriate to analyze $\mathcal{Q} \tilde{W'}^L$ with $L>1$.
The above analysis of the eigenvector space implies that the eigenvalues will be 
$\theta^L \Lambda_1^{[N^2-2N+1]}$, $\theta^L \Lambda_2^{[N-1]}$, $\Lambda_3^{[N-1]}$, and $\Lambda_4^{[1]}$: The ground state and the first excited state do not change, whereas the excitation to higher energy levels becomes less and less probable as the timescale difference $L$ grows if $\theta<1$. It reproduces the case of uniform weights in the limit of $L \to \infty$.

\subsubsection{Numerical simulations}
\label{sec:recovery_num}
Let us check if our analysis is corroborated by numerical simulations. We have conducted agent-based simulations, in which we consider a population of $N$ individuals who use a common norm, i.e., $\alpha_i = \alpha$ and $\beta_i = \beta$. At each time step, we randomly pick up a pair of individuals $i$ and $j$ for the donation game, where $i$ is the donor and $j$ is the recipient. 
The donor calculates the cooperation level $\beta (m_{ii}^t, m_{ij}^t)$, so that the donor pays $c\beta (m_{ii}^t, m_{ij}^t)$ and the recipient earns $b\beta (m_{ii}^t, m_{ij}^t)$. The interaction between $i$ and $j$ is observed by another player $k$ with probability $q$, while the donor and the recipient observe their own interaction with certainty. The observer $k$ updates his or her own opinion about the donor by using the assessment rule $\alpha \left[ m_{ki}^t, \beta (m_{ii}^t, m_{ij}^t), m_{kj}^t \right]$. If $q \sim O(1)$, each interaction changes $O(N)$ elements of the image matrix. Hereafter, $N$ interactions will be called one Monte Carlo step (MCS).

\begin{figure}
\includegraphics[width=0.45\textwidth]{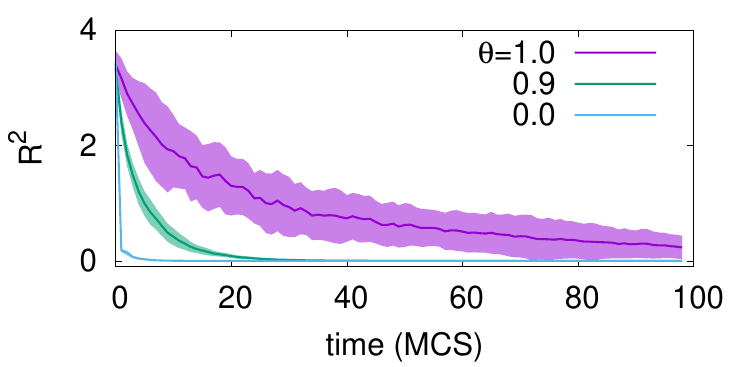}
\caption{The effect of $\theta$ on the high-order excitations of the normative dynamics. We consider $N=50$ individuals who use Simple Standing (L3). The initial condition is close to $m^\ast=1$, except that we introduce perturbation by randomly choosing $20\%$ of the image matrix elements and changing the values to $0.9$. The observation probability is $q=0.5$. The shades mean standard error estimated from $10$ independent runs.}
\label{fig:remain}
\end{figure}

The simulation keeps track of the time evolution of $\vec{m}^t = (m_{11}^t, \ldots, m_{NN}^t)$ in an $N^2$-dimensional space $V$. Let $Z$ denote a subspace of $V$, spanned by the eigenvectors in Eq.~\eqref{eq:uk} that correspond to the eigenvalue $\mu_1^{[N(N-1)]}=1$. The projection of $\vec{m}^t$ onto $Z$ will be denoted by $\vec{m}^t_Z$, whereas the remaining part in $V\setminus Z$ is denoted by
\begin{equation}
\vec{R} \equiv \vec{m}^t - \vec{m}^t_Z.
\label{eq:rem}
\end{equation}
After updating $\vec{m}^t$ by 1 MCS, we multiply it by $\tilde{W}' (\theta)$ for aggregating private opinions. 
Figure~\ref{fig:remain} depicts how $R^2 \equiv \vec{R}\cdot\vec{R}$ decreases with time for different values of $\theta$. The case of $\theta=1$ is shown for comparison because $W'(\theta=1)$ is an identity matrix, whereby no opinion aggregation occurs. We can clearly see that high-order excitations in $V \setminus Z$ are suppressed when $\theta<1$, consistently with our analytic prediction.

\subsection{Invasion analysis}
\subsubsection{Without opinion aggregation}

Now we wish to analyze the invasion of mutants.
Let us begin by reviewing the existing results~\cite{lee2021local}. We will divide the whole population into two groups: One consists of mutants, indexed by $0$, and it constitutes a small fraction $p$ of the population. The other individuals are residents with index $1$.
The resident population all have an assessment rule $\alpha_1$ and an behavioral rule $\beta_1$. The mutant has $\alpha_0$ and $\beta_0$, which are different from the residents' by small amounts, i.e., $\alpha_0(x,y,z) = \alpha_1(x,y,z) - \delta(x,y,z)$ and $\beta_0(x,y) = \beta_1(x,y) - \eta(x,y)$ with $|\delta| \ll 1$ and $|\eta| \ll 1$.
By setting $q=1$ in Eq.~\eqref{eq:update} for the sake of simplicity,
we can derive a set of equations as follows:
\begin{subequations}
\begin{align}
m_{00}^{t+1} &= p \alpha_0 [m_{00}^t, \beta_0(m_{00}^t, m_{00}^t), m_{00}^t] + \bar{p}
\alpha_0 [m_{00}^t, \beta_0(m_{00}^t, m_{01}^t), m_{01}^t]\\
m_{01}^{t+1} &= p \alpha_0 [m_{01}^t, \beta_1(m_{11}^t, m_{10}^t), m_{00}^t] + \bar{p}
\alpha_0 [m_{01}^t, \beta_1(m_{11}^t, m_{11}^t), m_{01}^t]\\
m_{10}^{t+1} &= p \alpha_1 [m_{10}^t, \beta_0(m_{00}^t, m_{00}^t), m_{10}^t] + \bar{p}
\alpha_1 [m_{10}^t, \beta_0(m_{00}^t, m_{01}^t), m_{11}^t]\\
m_{11}^{t+1} &= p \alpha_1 [m_{11}^t, \beta_1(m_{11}^t, m_{10}^t), m_{10}^t] + \bar{p}
\alpha_1 [m_{11}^t, \beta_1(m_{11}^t, m_{11}^t), m_{11}^t],
\end{align}
\label{eq:mutant}
\end{subequations}
where $\bar{p} \equiv 1-p$.
The resident norm composed of $\alpha_1$ and $\beta_1$ is one of L1, L3, and L4, so that $m^\ast=1$. We assume $m_{00}^t = 1-\epsilon_{00}^t$, $m_{01}^t = 1-\epsilon_{01}^t$, $m_{10}^t = 1-\epsilon_{10}^t$, and $m_{11}^t = 1-\epsilon_{11}^t$ with $0 < \epsilon_{ij}^t \ll 1$, and expand Eq.~\eqref{eq:mutant} to the first order of $\epsilon_{ij}^t$'s as has been done in Eq.~\eqref{eq:disagree} to obtain
\begin{subequations}
\begin{align}
1-\epsilon_{00}^{t+1} &\approx p [1-A_x \epsilon_{00}^t - A_y (B_x
\epsilon_{00}^t + B_y \epsilon_{00}^t + \eta_1) - A_z \epsilon_{00}^t -
\delta_1]\nonumber\\
&+ \bar{p} [1-A_x \epsilon_{00}^t - A_y (B_x \epsilon_{00}^t +
B_y \epsilon_{01}^t + \eta_1) - A_z \epsilon_{01}^t - \delta_1]\\
1-\epsilon_{01}^{t+1} &\approx p [1-A_x \epsilon_{01}^t - A_y (B_x
\epsilon_{11}^t + B_y \epsilon_{10}^t) - A_z \epsilon_{00}^t -
\delta_1]\nonumber\\
&+ \bar{p} [1-A_x \epsilon_{01}^t - A_y (B_x \epsilon_{11}^t +
B_y \epsilon_{11}^t) - A_z \epsilon_{01}^t - \delta_1]\\
1-\epsilon_{10}^{t+1} &\approx p [1-A_x \epsilon_{10}^t - A_y (B_x
\epsilon_{00}^t + B_y \epsilon_{00}^t + \eta_1) - A_z
\epsilon_{10}^t]\nonumber\\
&+ \bar{p} [1-A_x \epsilon_{10}^t - A_y (B_x \epsilon_{00}^t +
B_y \epsilon_{01}^t + \eta_1) - A_z \epsilon_{11}^t]\\
1-\epsilon_{11}^{t+1} &\approx p [1-A_x \epsilon_{11}^t - A_y (B_x
\epsilon_{11}^t + B_y \epsilon_{10}^t) - A_z
\epsilon_{10}^t]\nonumber\\
&+ \bar{p} [1-A_x \epsilon_{11}^t - A_y (B_x \epsilon_{11}^t +
B_y \epsilon_{11}^t) - A_z \epsilon_{11}^t].
\end{align}
\label{eq:linear}
\end{subequations}
In the long-time limit, the system will reach a stationary state at which $\epsilon_{ij}^\ast = \lim_{t\to \infty}\epsilon_{ij}^t$.
After some linear algebra, we obtain the following results in the small-$p$ limit:
\begin{subequations}
\begin{align}
\epsilon_{00}^\ast &= \frac{(1-A_x+A_y B_y)\delta_1 +
(1-A_x-A_z)A_y \eta_1}{(1-A_x-A_z)(1-A_x-A_y
B_x)}\label{eq:epsilon00}\\
\epsilon_{01}^\ast &= \frac{\delta_1}{1-A_x-A_z}\\
\epsilon_{10}^\ast &= \frac{(B_x+B_y)\delta_1 +
(1-A_x-A_z)\eta_1}{(1-A_x-A_z)(1-A_x-A_y B_x)}A_y\\
\epsilon_{11}^\ast &= 0,
\end{align}
\end{subequations}
under the condition that $A_x+A_z<1$ and $A_x+A_yB_x < 1$, where
$\delta_1 \equiv \delta(1,1,1) >0$ and $\eta_1 \equiv \eta(1,1) >0$.
In the donation game parameterized by benefit $b$ and cost $c$, a resident individual earns $\pi_1 = (1/2)(b-c)$ per interaction, where the factor of $1/2$ means that the individual can be either a donor or a recipient in an interaction, whereas the payoff earned by a mutant is
\begin{subequations}
\begin{align}
\pi_0 &= \frac{1}{2} \left[ b \beta_1 (m_{11}^\ast, m_{10}^\ast) -c \beta_0 (m_{00}^\ast, m_{01}^\ast) \right]\\
&\approx \frac{1}{2} \left[ b(1-B_y\epsilon_{10}^\ast) -c(1-B_x \epsilon_{00}^\ast-B_y \epsilon_{01}^\ast - \eta_1) \right].
\end{align}
\label{eq:pi0}
\end{subequations}
If we check when a mutant becomes worse off than a resident, i.e.,
$\Delta \pi \equiv \pi_0 - \pi_1<0$, the condition turns out to be
\begin{equation}
\frac{b}{c} > \frac{1-A_x}{A_yB_y},
\label{eq:bc_ratio}
\end{equation}
regardless of the value of $p$.
The right-hand side is unity for all the three norms that have $m^\ast=1$ as an asymptotically stable fixed point, i.e., L1, L3, and L4, so we conclude that they successfully suppress mutation in any donation game with $b>c$.

\subsubsection{With opinion aggregation}

Now, let us combine the process of reputation formation here. For the sake of simplicity, we assume that the averaging process uses uniform weights and has the same time scale as the normative dynamics. The important point is that an average opinion is dominated by the residents when $N \gg 1$ and $p\ll 1$ with keeping the number of mutants $Np$ constant. Mathematically speaking, every time when the system is updated by Eq.~\eqref{eq:linear}, we have to apply another linear operation to replace $\epsilon_{01}^t$ and $\epsilon_{00}^t$ by $\langle \epsilon_{01}^t \rangle \equiv p\epsilon_{01}^t + \bar{p}\epsilon_{11}^t$ and $\langle \epsilon_{00}^t \rangle \equiv p \epsilon_{00}^t + \bar{p} \epsilon_{10}^t$, respectively.
The stationary state of the combined dynamics is again obtained through linear algebra. The full solution is not very illuminating, so we write only its small-$p$ limit as follows:
\begin{subequations}
\begin{align}
{\epsilon_{00}^\ast}' &= \delta_1 + \frac{A_y}{1-A_x-A_y B_x}\eta_1\\
{\epsilon_{01}^\ast}' &= \delta_1\\
{\epsilon_{10}^\ast}' &= \frac{A_y}{1-A_x-A_y B_x}\eta_1\\
{\epsilon_{11}^\ast}' &= 0.
\end{align}
\end{subequations}
By substituting these parameters into Eq.~\eqref{eq:pi0}, we obtain
\begin{equation}
\pi_0' \approx \frac{1}{2} \left[ b(1-B_y{\langle \epsilon_{10}^\ast}' \rangle) -c(1-B_x \langle {\epsilon_{00}^\ast}' \rangle -B_y \langle {\epsilon_{01}^\ast}' \rangle - \eta_1) \right],
\end{equation}
where 
\begin{subequations}
\begin{align}
\langle {\epsilon_{10}^\ast}' \rangle &\equiv p{\epsilon_{00}^\ast}' + \bar{p} {\epsilon_{10}^\ast}' \xrightarrow[p\to 0]{}{\epsilon_{10}^\ast}'\\
\langle {\epsilon_{00}^\ast}' \rangle &\equiv p{\epsilon_{00}^\ast}' + \bar{p} {\epsilon_{10}^\ast}' \xrightarrow[p\to 0]{}{\epsilon_{10}^\ast}'\\
\langle {\epsilon_{01}^\ast}' \rangle &\equiv p{\epsilon_{01}^\ast}' + \bar{p} {\epsilon_{11}^\ast}' \xrightarrow[p\to 0]{}{\epsilon_{11}^\ast}'.
\end{align}
\end{subequations}
The difference of a mutant's average payoff from that of a resident player is now calculated to be
\begin{eqnarray}
\Delta \pi' &=&
- \frac{[bA_yB_y - c(1-A_x)]}{1-A_x-A_yB_x} \eta_1\nonumber\\
&-&p\frac{[bB_y(1-A_x+A_yB_y)-(1-A_x)(B_x+B_y)c]}{1-A_x-A_yB_x} \delta_1\nonumber\\
 &-&p \frac{A_y (A_z+A_yB_y) [bB_y -(B_x+B_y)c]}{(1-A_x-A_yB_x) (1-A_x-A_z-A_yB_x-A_yB_y)} \eta_1 + O(p^2).
 \label{eq:dpi'}
\end{eqnarray}
Thus, if we consider $\alpha$-mutants by setting $\eta_1=0$, Eq.~\eqref{eq:dpi'} vanishes in the limit of $p\to 0$, so that the $\alpha$-mutants can neutrally invade the population.
By contrast, by setting $\delta_1=0$, we can see that $\beta$-mutants have the same stability condition as given by Eq.~\eqref{eq:bc_ratio} when $p \ll 1$.

\subsubsection{Numerical simulations}

\begin{figure}
\centering
\includegraphics[width=0.45\textwidth]{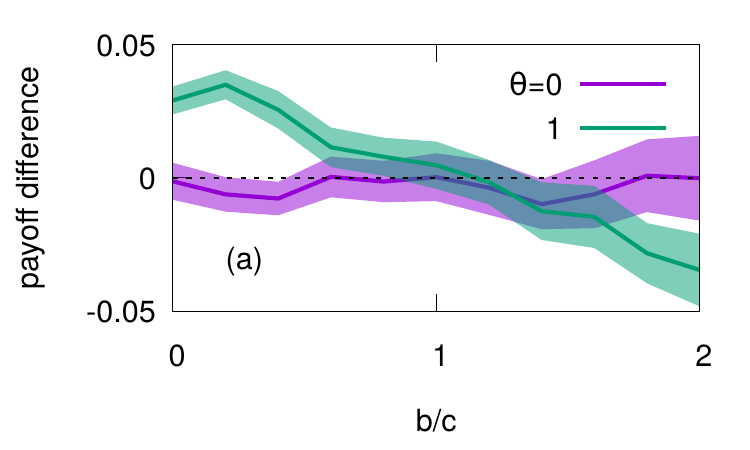}
\includegraphics[width=0.45\textwidth]{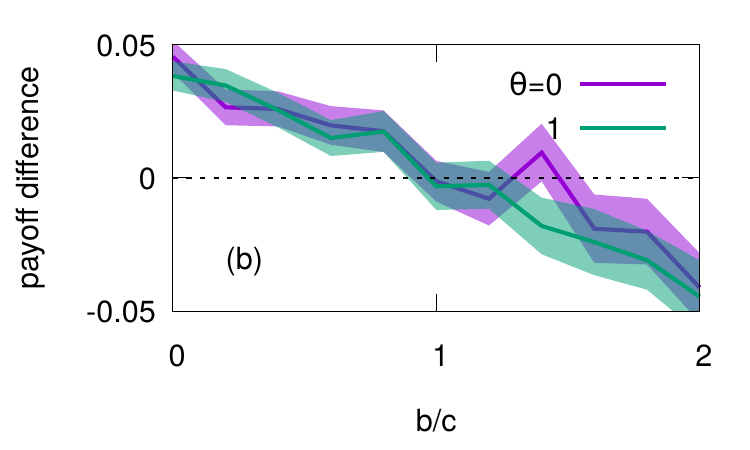}
\caption{Invasion analysis of a resident population using Simple Standing (L3). (a) The mutant norm has a small difference $\delta(x,y,z) = \delta_1 (2yz-2z+1)$ with $\delta_1=5 \times 10^{-2}$ from the residents' $\alpha(x,y,z) = yz-z+1$ (see Table~\ref{tab:cont}). (b) The mutant norm has a small difference $\eta(x,y) = \eta_1 xy$ with $\eta_1 = 5\times 10^{-2}$ from the residents' $\beta(x,y) = y$. The population size is $N=50$, the fraction of mutants is $p=0.1$, and the observation probability is $q=0.5$. We have generated $10^2$ independent samples for each data point to estimate standard error as represented by the shades. Each sample has been simulated for $10^2$ MCS, out of which the last half has been used for calculating the payoffs.}
\label{fig:inv}
\end{figure}

Again, we have conducted agent-based simulations of $N$ players, among which $Np$ mutants have a different social norm. If a player $k$ is a resident, the player uses $\alpha_1$ and $\beta_1$, whereas the player uses $\alpha_0 = \alpha_1 - \delta$  and $\beta_0 = \beta_1 - \eta$ if a mutant.
Randomly chosen players $i$ and $j$ play the donation game as a donor and a recipient, respectively, while being observed by the population as before. The difference from Sec.~\ref{sec:recovery_num} is that the donor's behavior and the observers' assessments are based on average opinions instead of private ones: That is, initially we begin with $m_{ki}^{t=0} = 1$ for every pair of $k$ and $i$. As a result of aggregation, we thus have $m_l^{t=0} \equiv N^{-1} \sum_k m_{kl}^{t=0} = 1$.
During the first MCS, the image matrix is updated through $N$ rounds of interaction, and the aggregated opinions are updated accordingly. At the next MCS, donor $i$'s cooperation level is given by $\beta_i (m_i^{t=0}, m_j^{t=0})$, and each observer's assessment is updated by using $\alpha_k [m_i^{t=0}, \beta_i (m_i^{t=0}, m_j^{t=0}), m_j^{t=0}]$. In this way, the aggregation and normative dynamics are repeated alternately until reaching a steady state. At this steady state, we calculate the payoff difference between a resident player and a mutant.

Figure~\ref{fig:inv}(a) shows that the averaging process with uniform weights ($\theta=0$) weakens the stability against $\alpha$-mutants when the resident norm is Simple Standing (L3): In the absence of opinion aggregation ($\theta=1$), as long as $b/c>1$, the resident norm is able to incur a payoff cost to any mutant norm with small $\delta$. It is no longer the case when $\theta=0$, for which the payoff difference is statistically zero. As for $\beta$-mutation, the shape of the curve does not change qualitatively whether private opinions are aggregated or not [Fig.~\ref{fig:inv}(b)]. All those numerical results support the validity of our analysis.

\section{Discussion and Summary}
\label{sec:discuss}

In summary, we have argued that reputation, understood as a publicly shared opinion, should be considered separately from assessment, which is basically private. We have proposed a mathematical framework for reputation formation represented by a doubly stochastic matrix. The averaging process described by $W'(\theta)$ is a special case which, when combined with the continuous model of indirect reciprocity, lends itself to mathematical analysis because of its simple eigenvalue structure.
The analysis shows that our combined dynamics preserves the stability of the original normative dynamics to a large extent, and it is because both are based on private assessments, whether aggregated or not. Yet, the stability against $\alpha$-mutants is altered by the averaging process.

Even if we relax the bi-stochasticity to stochasticity, the convergence to the principal eigenvector is still guaranteed by the Perron-Frobenius theorem. The elements of the principal eigenvector are no longer uniform but only have the same sign. Therefore, any two individuals $i$ and $j$ have different opinions about others, i.e., $\left( m_{i1}, m_{i2}, \ldots, m_{iN} \right) \neq \left( m_{j1}, m_{j2}, \ldots, m_{jN} \right)$ in general, but these two vectors differ only by a constant factor when the system has converged to the principal eigenvector. In this sense, weighted average can be a powerful and ubiquitous mechanism for reaching a consensus, as long as we neglect the proportionality constant in front of each individual's opinion vector.

We have relied on linear-algebraic operations to obtain reputation as a consensus of opinions.
However, it is intuitively plausible that opinion aggregation in practice may contain non-linearity, e.g., by having the weight $w_{ij}$ correlated with $m_{ij}$ itself. Then, this non-linearity may well break the consensus~\cite{marvel2011continuous}, making someone highly respected among a group of people but regarded as a villain by the rest.
Although reputation does not exist here in the strict sense, such a ``cult'' is an intriguing aspect of human interaction involved in indirect reciprocity, which calls for further development of our framework.

\section*{Funding}
We acknowledge support by Basic Science Research Program through the
National Research Foundation of Korea (NRF) funded by the Ministry of Education
(NRF-2020R1I1A2071670).

\bibliographystyle{apsrev4-1}
\bibliography{main}

\begin{thebibliography}{24}%
\makeatletter
\providecommand \@ifxundefined [1]{%
 \@ifx{#1\undefined}
}%
\providecommand \@ifnum [1]{%
 \ifnum #1\expandafter \@firstoftwo
 \else \expandafter \@secondoftwo
 \fi
}%
\providecommand \@ifx [1]{%
 \ifx #1\expandafter \@firstoftwo
 \else \expandafter \@secondoftwo
 \fi
}%
\providecommand \natexlab [1]{#1}%
\providecommand \enquote  [1]{``#1''}%
\providecommand \bibnamefont  [1]{#1}%
\providecommand \bibfnamefont [1]{#1}%
\providecommand \citenamefont [1]{#1}%
\providecommand \href@noop [0]{\@secondoftwo}%
\providecommand \href [0]{\begingroup \@sanitize@url \@href}%
\providecommand \@href[1]{\@@startlink{#1}\@@href}%
\providecommand \@@href[1]{\endgroup#1\@@endlink}%
\providecommand \@sanitize@url [0]{\catcode `\\12\catcode `\$12\catcode `\&12\catcode `\#12\catcode `\^12\catcode `\_12\catcode `\%12\relax}%
\providecommand \@@startlink[1]{}%
\providecommand \@@endlink[0]{}%
\providecommand \url  [0]{\begingroup\@sanitize@url \@url }%
\providecommand \@url [1]{\endgroup\@href {#1}{\urlprefix }}%
\providecommand \urlprefix  [0]{URL }%
\providecommand \Eprint [0]{\href }%
\providecommand \doibase [0]{http://dx.doi.org/}%
\providecommand \selectlanguage [0]{\@gobble}%
\providecommand \bibinfo  [0]{\@secondoftwo}%
\providecommand \bibfield  [0]{\@secondoftwo}%
\providecommand \translation [1]{[#1]}%
\providecommand \BibitemOpen [0]{}%
\providecommand \bibitemStop [0]{}%
\providecommand \bibitemNoStop [0]{.\EOS\space}%
\providecommand \EOS [0]{\spacefactor3000\relax}%
\providecommand \BibitemShut  [1]{\csname bibitem#1\endcsname}%
\let\auto@bib@innerbib\@empty
\bibitem [{\citenamefont {Origgi}(2017)}]{origgi2017reputation}%
  \BibitemOpen
  \bibfield  {author} {\bibinfo {author} {\bibfnamefont {G.}~\bibnamefont {Origgi}},\ }\href@noop {} {\emph {\bibinfo {title} {Reputation: What it is and why it matters}}}\ (\bibinfo  {publisher} {Princeton University Press},\ \bibinfo {address} {Princeton, NJ},\ \bibinfo {year} {2017})\BibitemShut {NoStop}%
\bibitem [{\citenamefont {Dunbar}(2004)}]{dunbar2004gossip}%
  \BibitemOpen
  \bibfield  {author} {\bibinfo {author} {\bibfnamefont {R.~I.}\ \bibnamefont {Dunbar}},\ }\href@noop {} {\bibfield  {journal} {\bibinfo  {journal} {Rev. Gen. Psychol.}\ }\textbf {\bibinfo {volume} {8}},\ \bibinfo {pages} {100} (\bibinfo {year} {2004})}\BibitemShut {NoStop}%
\bibitem [{\citenamefont {Nowak}\ and\ \citenamefont {Sigmund}(2005)}]{nowak2005evolution}%
  \BibitemOpen
  \bibfield  {author} {\bibinfo {author} {\bibfnamefont {M.~A.}\ \bibnamefont {Nowak}}\ and\ \bibinfo {author} {\bibfnamefont {K.}~\bibnamefont {Sigmund}},\ }\href@noop {} {\bibfield  {journal} {\bibinfo  {journal} {Nature}\ }\textbf {\bibinfo {volume} {437}},\ \bibinfo {pages} {1291} (\bibinfo {year} {2005})}\BibitemShut {NoStop}%
\bibitem [{\citenamefont {Xia}\ \emph {et~al.}(2023)\citenamefont {Xia}, \citenamefont {Wang}, \citenamefont {Perc},\ and\ \citenamefont {Wang}}]{xia2023reputation}%
  \BibitemOpen
  \bibfield  {author} {\bibinfo {author} {\bibfnamefont {C.}~\bibnamefont {Xia}}, \bibinfo {author} {\bibfnamefont {J.}~\bibnamefont {Wang}}, \bibinfo {author} {\bibfnamefont {M.}~\bibnamefont {Perc}}, \ and\ \bibinfo {author} {\bibfnamefont {Z.}~\bibnamefont {Wang}},\ }\href@noop {} {\bibfield  {journal} {\bibinfo  {journal} {Phys. Life Rev.}\ }\textbf {\bibinfo {volume} {46}},\ \bibinfo {pages} {8} (\bibinfo {year} {2023})}\BibitemShut {NoStop}%
\bibitem [{\citenamefont {Santos}(2023)}]{santos2023consensus}%
  \BibitemOpen
  \bibfield  {author} {\bibinfo {author} {\bibfnamefont {F.~P.}\ \bibnamefont {Santos}},\ }\href@noop {} {\bibfield  {journal} {\bibinfo  {journal} {Phys. Life Rev.}\ }\textbf {\bibinfo {volume} {46}},\ \bibinfo {pages} {187} (\bibinfo {year} {2023})}\BibitemShut {NoStop}%
\bibitem [{\citenamefont {Nowak}\ and\ \citenamefont {Sigmund}(1998)}]{nowak1998evolution}%
  \BibitemOpen
  \bibfield  {author} {\bibinfo {author} {\bibfnamefont {M.~A.}\ \bibnamefont {Nowak}}\ and\ \bibinfo {author} {\bibfnamefont {K.}~\bibnamefont {Sigmund}},\ }\href@noop {} {\bibfield  {journal} {\bibinfo  {journal} {Nature}\ }\textbf {\bibinfo {volume} {393}},\ \bibinfo {pages} {573} (\bibinfo {year} {1998})}\BibitemShut {NoStop}%
\bibitem [{\citenamefont {Ohtsuki}\ and\ \citenamefont {Iwasa}(2004)}]{ohtsuki2004should}%
  \BibitemOpen
  \bibfield  {author} {\bibinfo {author} {\bibfnamefont {H.}~\bibnamefont {Ohtsuki}}\ and\ \bibinfo {author} {\bibfnamefont {Y.}~\bibnamefont {Iwasa}},\ }\href@noop {} {\bibfield  {journal} {\bibinfo  {journal} {J. Theor. Biol.}\ }\textbf {\bibinfo {volume} {231}},\ \bibinfo {pages} {107} (\bibinfo {year} {2004})}\BibitemShut {NoStop}%
\bibitem [{\citenamefont {Burke}(1985)}]{burke1985day}%
  \BibitemOpen
  \bibfield  {author} {\bibinfo {author} {\bibfnamefont {J.}~\bibnamefont {Burke}},\ }\href@noop {} {\emph {\bibinfo {title} {The Day the Universe Changed}}}\ (\bibinfo  {publisher} {London Writers Ltd.},\ \bibinfo {address} {London},\ \bibinfo {year} {1985})\BibitemShut {NoStop}%
\bibitem [{\citenamefont {Hilbe}\ \emph {et~al.}(2018)\citenamefont {Hilbe}, \citenamefont {Schmid}, \citenamefont {Tkadlec}, \citenamefont {Chatterjee},\ and\ \citenamefont {Nowak}}]{hilbe2018indirect}%
  \BibitemOpen
  \bibfield  {author} {\bibinfo {author} {\bibfnamefont {C.}~\bibnamefont {Hilbe}}, \bibinfo {author} {\bibfnamefont {L.}~\bibnamefont {Schmid}}, \bibinfo {author} {\bibfnamefont {J.}~\bibnamefont {Tkadlec}}, \bibinfo {author} {\bibfnamefont {K.}~\bibnamefont {Chatterjee}}, \ and\ \bibinfo {author} {\bibfnamefont {M.~A.}\ \bibnamefont {Nowak}},\ }\href@noop {} {\bibfield  {journal} {\bibinfo  {journal} {Proc. Natl. Acad. Sci. USA}\ }\textbf {\bibinfo {volume} {115}},\ \bibinfo {pages} {12241} (\bibinfo {year} {2018})}\BibitemShut {NoStop}%
\bibitem [{\citenamefont {Schmid}\ \emph {et~al.}(2021)\citenamefont {Schmid}, \citenamefont {Shati}, \citenamefont {Hilbe},\ and\ \citenamefont {Chatterjee}}]{schmid2021evolution}%
  \BibitemOpen
  \bibfield  {author} {\bibinfo {author} {\bibfnamefont {L.}~\bibnamefont {Schmid}}, \bibinfo {author} {\bibfnamefont {P.}~\bibnamefont {Shati}}, \bibinfo {author} {\bibfnamefont {C.}~\bibnamefont {Hilbe}}, \ and\ \bibinfo {author} {\bibfnamefont {K.}~\bibnamefont {Chatterjee}},\ }\href@noop {} {\bibfield  {journal} {\bibinfo  {journal} {Sci. Rep.}\ }\textbf {\bibinfo {volume} {11}},\ \bibinfo {pages} {17443} (\bibinfo {year} {2021})}\BibitemShut {NoStop}%
\bibitem [{\citenamefont {Schmid}\ \emph {et~al.}(2023)\citenamefont {Schmid}, \citenamefont {Ekbatani}, \citenamefont {Hilbe},\ and\ \citenamefont {Chatterjee}}]{schmid2023quantitative}%
  \BibitemOpen
  \bibfield  {author} {\bibinfo {author} {\bibfnamefont {L.}~\bibnamefont {Schmid}}, \bibinfo {author} {\bibfnamefont {F.}~\bibnamefont {Ekbatani}}, \bibinfo {author} {\bibfnamefont {C.}~\bibnamefont {Hilbe}}, \ and\ \bibinfo {author} {\bibfnamefont {K.}~\bibnamefont {Chatterjee}},\ }\href@noop {} {\bibfield  {journal} {\bibinfo  {journal} {Nat. Commun.}\ }\textbf {\bibinfo {volume} {14}},\ \bibinfo {pages} {2086} (\bibinfo {year} {2023})}\BibitemShut {NoStop}%
\bibitem [{\citenamefont {Fujimoto}\ and\ \citenamefont {Ohtsuki}(2023)}]{fujimoto2023evolutionary}%
  \BibitemOpen
  \bibfield  {author} {\bibinfo {author} {\bibfnamefont {Y.}~\bibnamefont {Fujimoto}}\ and\ \bibinfo {author} {\bibfnamefont {H.}~\bibnamefont {Ohtsuki}},\ }\href@noop {} {\bibfield  {journal} {\bibinfo  {journal} {Proc. Natl. Acad. Sci. USA}\ }\textbf {\bibinfo {volume} {120}},\ \bibinfo {pages} {e2300544120} (\bibinfo {year} {2023})}\BibitemShut {NoStop}%
\bibitem [{\citenamefont {Murase}\ and\ \citenamefont {Hilbe}(2024{\natexlab{a}})}]{murase2024computational}%
  \BibitemOpen
  \bibfield  {author} {\bibinfo {author} {\bibfnamefont {Y.}~\bibnamefont {Murase}}\ and\ \bibinfo {author} {\bibfnamefont {C.}~\bibnamefont {Hilbe}},\ }\href@noop {} {\bibfield  {journal} {\bibinfo  {journal} {Proc. Natl. Acad. Sci. USA}\ }\textbf {\bibinfo {volume} {121}},\ \bibinfo {pages} {e2406885121} (\bibinfo {year} {2024}{\natexlab{a}})}\BibitemShut {NoStop}%
\bibitem [{\citenamefont {Ohtsuki}\ \emph {et~al.}(2009)\citenamefont {Ohtsuki}, \citenamefont {Iwasa},\ and\ \citenamefont {Nowak}}]{ohtsuki2009indirect}%
  \BibitemOpen
  \bibfield  {author} {\bibinfo {author} {\bibfnamefont {H.}~\bibnamefont {Ohtsuki}}, \bibinfo {author} {\bibfnamefont {Y.}~\bibnamefont {Iwasa}}, \ and\ \bibinfo {author} {\bibfnamefont {M.~A.}\ \bibnamefont {Nowak}},\ }\href@noop {} {\bibfield  {journal} {\bibinfo  {journal} {Nature}\ }\textbf {\bibinfo {volume} {457}},\ \bibinfo {pages} {79} (\bibinfo {year} {2009})}\BibitemShut {NoStop}%
\bibitem [{\citenamefont {Tak{\'a}cs}\ \emph {et~al.}(2021)\citenamefont {Tak{\'a}cs}, \citenamefont {Gross}, \citenamefont {Testori}, \citenamefont {Letina}, \citenamefont {Kenny}, \citenamefont {Power},\ and\ \citenamefont {Wittek}}]{takacs2021networks}%
  \BibitemOpen
  \bibfield  {author} {\bibinfo {author} {\bibfnamefont {K.}~\bibnamefont {Tak{\'a}cs}}, \bibinfo {author} {\bibfnamefont {J.}~\bibnamefont {Gross}}, \bibinfo {author} {\bibfnamefont {M.}~\bibnamefont {Testori}}, \bibinfo {author} {\bibfnamefont {S.}~\bibnamefont {Letina}}, \bibinfo {author} {\bibfnamefont {A.~R.}\ \bibnamefont {Kenny}}, \bibinfo {author} {\bibfnamefont {E.~A.}\ \bibnamefont {Power}}, \ and\ \bibinfo {author} {\bibfnamefont {R.~P.}\ \bibnamefont {Wittek}},\ }\href@noop {} {\bibfield  {journal} {\bibinfo  {journal} {Philos. Trans. R. Soc. B}\ }\textbf {\bibinfo {volume} {376}},\ \bibinfo {pages} {20200297} (\bibinfo {year} {2021})}\BibitemShut {NoStop}%
\bibitem [{\citenamefont {Kessinger}\ \emph {et~al.}(2023)\citenamefont {Kessinger}, \citenamefont {Tarnita},\ and\ \citenamefont {Plotkin}}]{kessinger2023evolution}%
  \BibitemOpen
  \bibfield  {author} {\bibinfo {author} {\bibfnamefont {T.~A.}\ \bibnamefont {Kessinger}}, \bibinfo {author} {\bibfnamefont {C.~E.}\ \bibnamefont {Tarnita}}, \ and\ \bibinfo {author} {\bibfnamefont {J.~B.}\ \bibnamefont {Plotkin}},\ }\href@noop {} {\bibfield  {journal} {\bibinfo  {journal} {Proc. Natl. Acad. Sci. USA}\ }\textbf {\bibinfo {volume} {120}},\ \bibinfo {pages} {e2219480120} (\bibinfo {year} {2023})}\BibitemShut {NoStop}%
\bibitem [{\citenamefont {Kawakatsu}\ \emph {et~al.}(2024)\citenamefont {Kawakatsu}, \citenamefont {Kessinger},\ and\ \citenamefont {Plotkin}}]{kawakatsu2024mechanistic}%
  \BibitemOpen
  \bibfield  {author} {\bibinfo {author} {\bibfnamefont {M.}~\bibnamefont {Kawakatsu}}, \bibinfo {author} {\bibfnamefont {T.~A.}\ \bibnamefont {Kessinger}}, \ and\ \bibinfo {author} {\bibfnamefont {J.~B.}\ \bibnamefont {Plotkin}},\ }\href@noop {} {\bibfield  {journal} {\bibinfo  {journal} {Proc. Natl. Acad. Sci. USA}\ }\textbf {\bibinfo {volume} {121}},\ \bibinfo {pages} {e2400689121} (\bibinfo {year} {2024})}\BibitemShut {NoStop}%
\bibitem [{\citenamefont {Murase}\ and\ \citenamefont {Hilbe}(2024{\natexlab{b}})}]{murase2024indirect}%
  \BibitemOpen
  \bibfield  {author} {\bibinfo {author} {\bibfnamefont {Y.}~\bibnamefont {Murase}}\ and\ \bibinfo {author} {\bibfnamefont {C.}~\bibnamefont {Hilbe}},\ }\href@noop {} {\enquote {\bibinfo {title} {Indirect reciprocity under opinion synchronization},}\ }\bibinfo {howpublished} {arXiv preprint arXiv:2409.05551} (\bibinfo {year} {2024}{\natexlab{b}})\BibitemShut {NoStop}%
\bibitem [{\citenamefont {Lee}\ \emph {et~al.}(2021)\citenamefont {Lee}, \citenamefont {Murase},\ and\ \citenamefont {Baek}}]{lee2021local}%
  \BibitemOpen
  \bibfield  {author} {\bibinfo {author} {\bibfnamefont {S.}~\bibnamefont {Lee}}, \bibinfo {author} {\bibfnamefont {Y.}~\bibnamefont {Murase}}, \ and\ \bibinfo {author} {\bibfnamefont {S.~K.}\ \bibnamefont {Baek}},\ }\href@noop {} {\bibfield  {journal} {\bibinfo  {journal} {Sci. Rep.}\ }\textbf {\bibinfo {volume} {11}},\ \bibinfo {pages} {14225} (\bibinfo {year} {2021})}\BibitemShut {NoStop}%
\bibitem [{\citenamefont {Lee}\ \emph {et~al.}(2022)\citenamefont {Lee}, \citenamefont {Murase},\ and\ \citenamefont {Baek}}]{lee2022second}%
  \BibitemOpen
  \bibfield  {author} {\bibinfo {author} {\bibfnamefont {S.}~\bibnamefont {Lee}}, \bibinfo {author} {\bibfnamefont {Y.}~\bibnamefont {Murase}}, \ and\ \bibinfo {author} {\bibfnamefont {S.~K.}\ \bibnamefont {Baek}},\ }\href@noop {} {\bibfield  {journal} {\bibinfo  {journal} {J. Theor. Biol.}\ }\textbf {\bibinfo {volume} {548}},\ \bibinfo {pages} {111202} (\bibinfo {year} {2022})}\BibitemShut {NoStop}%
\bibitem [{\citenamefont {Mun}\ and\ \citenamefont {Baek}(2024)}]{mun2023second}%
  \BibitemOpen
  \bibfield  {author} {\bibinfo {author} {\bibfnamefont {Y.}~\bibnamefont {Mun}}\ and\ \bibinfo {author} {\bibfnamefont {S.~K.}\ \bibnamefont {Baek}},\ }\href@noop {} {\bibfield  {journal} {\bibinfo  {journal} {Eur. Phys. J. Spec. Top.}\ }\textbf {\bibinfo {volume} {233}},\ \bibinfo {pages} {1251} (\bibinfo {year} {2024})}\BibitemShut {NoStop}%
\bibitem [{\citenamefont {Fujimoto}\ and\ \citenamefont {Ohtsuki}(2024)}]{fujimoto2024leader}%
  \BibitemOpen
  \bibfield  {author} {\bibinfo {author} {\bibfnamefont {Y.}~\bibnamefont {Fujimoto}}\ and\ \bibinfo {author} {\bibfnamefont {H.}~\bibnamefont {Ohtsuki}},\ }\href@noop {} {\bibfield  {journal} {\bibinfo  {journal} {PRX Life}\ }\textbf {\bibinfo {volume} {2}},\ \bibinfo {pages} {023009} (\bibinfo {year} {2024})}\BibitemShut {NoStop}%
\bibitem [{\citenamefont {Ohtsuki}\ and\ \citenamefont {Iwasa}(2006)}]{ohtsuki2006leading}%
  \BibitemOpen
  \bibfield  {author} {\bibinfo {author} {\bibfnamefont {H.}~\bibnamefont {Ohtsuki}}\ and\ \bibinfo {author} {\bibfnamefont {Y.}~\bibnamefont {Iwasa}},\ }\href@noop {} {\bibfield  {journal} {\bibinfo  {journal} {J. Theor. Biol.}\ }\textbf {\bibinfo {volume} {239}},\ \bibinfo {pages} {435} (\bibinfo {year} {2006})}\BibitemShut {NoStop}%
\bibitem [{\citenamefont {Marvel}\ \emph {et~al.}(2011)\citenamefont {Marvel}, \citenamefont {Kleinberg}, \citenamefont {Kleinberg},\ and\ \citenamefont {Strogatz}}]{marvel2011continuous}%
  \BibitemOpen
  \bibfield  {author} {\bibinfo {author} {\bibfnamefont {S.~A.}\ \bibnamefont {Marvel}}, \bibinfo {author} {\bibfnamefont {J.}~\bibnamefont {Kleinberg}}, \bibinfo {author} {\bibfnamefont {R.~D.}\ \bibnamefont {Kleinberg}}, \ and\ \bibinfo {author} {\bibfnamefont {S.~H.}\ \bibnamefont {Strogatz}},\ }\href@noop {} {\bibfield  {journal} {\bibinfo  {journal} {Proc. Natl. Acad. Sci. USA}\ }\textbf {\bibinfo {volume} {108}},\ \bibinfo {pages} {1771} (\bibinfo {year} {2011})}\BibitemShut {NoStop}%
\end{thebibliography}%

\end{document}